\documentclass[conference]{IEEEtran}
\IEEEoverridecommandlockouts

\usepackage{graphicx}
\usepackage{amsmath}
\usepackage{amssymb}
\usepackage{mathtools}
\usepackage{hyperref}
\usepackage{cleveref}
\usepackage{algorithm}
\usepackage{blindtext}
\usepackage{algpseudocode}
\usepackage{listings}
\usepackage{booktabs}
\usepackage{multirow}
\usepackage{subfig}
\usepackage{float}
\usepackage{multicol}
\usepackage{tikz}
\usepackage{cite}
\usepackage{url}
\usetikzlibrary{arrows.meta}
\usepackage{adjustbox}
\usepackage{authblk}
\usepackage[binary-units=true]{siunitx}

\newcommand{\mcimone}{Monte Cimone}

\begin{document}
\title{Monte Cimone: Paving the Road for the First Generation of RISC-V High-Performance Computers}

\author{
Andrea Bartolini,
Federico Ficarelli,
Emanuele Parisi,
Francesco Beneventi,
Francesco Barchi,
Daniele Gregori,\\
Fabrizio Magugliani,
Marco Cicala,
Cosimo Gianfreda,
Daniele Cesarini,
Andrea Acquaviva,
and~Luca Benini 

\IEEEcompsocitemizethanks{
    \IEEEcompsocthanksitem A. Bartolini, E. Parisi, F. Beneventi F. Barchi, A. Acquaviva, and L. Benini are with the Department of Electrical, Electronic and Information Engineering "Guglielmo Marconi", University of Bologna, 40136 Bologna, Italy (e-mail: a.bartolini@unibo.it; emanuele.parisi@unibo.it; francesco.beneventi@unibo.it; francesco.barchi@unibo.it; andrea.acquaviva@unibo.it; luca.benini@unibo.it ).
    \IEEEcompsocthanksitem F. Ficarelli and D. Cesarini are with the Department of SuperComputing Applications and Innovation, CINECA, 40033 Casalecchio di Reno (BO), Italy (e-mail: d.cesarini@cineca.it; f.ficarelli@cineca.it).
    \IEEEcompsocthanksitem D. Gregori, F. Magugliani, D. Cicala, and C. Gianfreda are with E4 Computer Engineering, 42019 Scandiano (RE), Italy (e-mail: daniele.gregori@e4company.com; fabrizio.maguglianie4company.com; marco.cicala@e4company.com; cosimo.gianfreda@e4company.com).}
}

\maketitle

\begin{abstract} 
The new open and royalty-free RISC-V ISA is attracting interest across the whole computing continuum, from microcontrollers to supercomputers. High-performance RISC-V processors and accelerators have been announced, but RISC-V-based HPC systems will need a holistic co-design effort, spanning memory, storage hierarchy interconnects and full software stack.  In this paper, we describe Monte Cimone, a fully-operational multi-blade computer prototype and hardware-software test-bed based on  U740, a double-precision capable multi-core, 64-bit RISC-V SoC. Monte Cimone does not aim to achieve strong floating-point performance, but it was built with the purpose of "priming the pipe" and exploring the challenges of integrating a multi-node RISC-V cluster capable of providing an HPC production stack including interconnect, storage and power monitoring infrastructure on RISC-V hardware. We present the results of our hardware/software integration effort, which demonstrate a remarkable level of software and hardware readiness and maturity - showing that the first generation of RISC-V HPC machines may not be so far in the future.
\end{abstract}

\section{Introduction}  \label{sec:intro}

The strategic role of High Performance Computing (HPC) systems is widely acknowledged in many fields, from weather forecasting to drug design. With the pervasive digitalization of our society, high performance computers fuel the most disruptive mega-trends, from the deployment of artificial intelligence (AI) at scale (e.g. for training large machine learning models) to industrial internet-of-things (IoT) applications (e.g. for creating and maintaining digital twins). Thus, HPC systems are today strategic assets not only for academia and industry, but also as for public institutions and governments\cite{SRA4}. 

The key challenge in designing HPC systems today and in the foreseable future is increasing compute efficiency, to meet the rapidly growing performance demand (10x every four years) within a constant or modestly increasing power budget, while facing the slow-down of Moore's Law. To exacerbate the efficiency challenge, while integrated circuits technology is still delivering device density increases (albeit as a slower pace), power consumption does not scale down at the same rate. Hence power density grows and it is increasingly difficult to meet thermal design power specifications without compromising performance. Disruptive technologies, such as quantum or optical computing  may bring long-term relief in some specific application areas, but there is no silver bullet in sight. 

To tackle the efficiency issue, academia and industry are aggressively pursuing architectural innovation and co-design strategies to develop HPC systems that mitigate the efficiency limitations of programmable architectures through various forms of specialization and domain-specific adaptation. Instruction Set Architectures (ISAs) have to evolve rapidly to sustain architectural evolution and domain adaptation, and the advent of the RISC-V open, royalty-free and extensible ISA has been a major step toward accelerating innovation in this area. An additional advantage of RISC-V with respect to the dominant proprietary ISAs (x86 and ARM) is that it is maintained by a global non-for-profit foundation with members across the world, ensuring a high degree of neutrality with respect to geopolitical tensions and their technology downfalls. 

Currently, high-performance 64bit (RV64) RISC-V processors and accelerator chips are being designed, promising prototypes are demonstrated in numerous publications \cite{hennessy2019golden} and products are announced at a fast cadence \cite{Dorflinger2021comparative, riscvxchange}. It is thus reasonable to expect that high-performance chips based on RISC-V will be available as production silicon within the next couple of years.  However, building a HPC system requires significantly more than just high-performance chips. Many think that the RISC-V software stack and system platform are extremely immature, and will need several additional years of development effort before full applications could be run, benchmarked and optimized on a RISC-V-based HPC system. Our goal is dispel this overly conservative notion. 

The main contribution of this work is to present \mcimone, the first physical prototype and test-bed of a complete RISC-V (RV64) compute cluster, integrating not only all the key hardware elements besides processors, namely main memory, non-volatile storage and interconnect, but also a complete software environment for HPC, as well as a full-featured system monitoring infrastructure. Further, we demonstrate that it is possible to run real-life HPC applications on \mcimone~today. Even though achieving strong double precision performance will be possible only with upcoming high-performance chips, we achieved the following milestones: 
\begin{itemize}
    \item We designed and set up the first RISC-V-based cluster containing eight computing nodes enclosed in four computing blades. Each computing node is based on the U740 SoC from SiFive and integrates four U74 RV64GCB application cores, running up to 1.2 GHz and 16GB of DDR4, 1 TB node-local NVME storage, and PCIe expansion cards. The cluster is connected to a login node and master node running the job scheduler, network file system and system management software. 
    \item We ported and assessed the maturity of a HPC software stack composed of (i) SLURM job scheduler, NAS filesystem, LDAP server, Spack package manager (ii) compilers toolchains, scientific and communication libraries, (iii) a set of HPC benchmarks  and applications, (iv) the ExaMon datacenter automation and monitoring framework. 
    \item We measured the efficiency of the HPL benchmark and STREAM benchmark with the toolchain and libraries installed by the SPACK. 
    We compared the attained results 
    against the one obtained for other RISC ISA architectures used in the 1st and 2nd ranked Top500 supercomputers (namely, Summit and Fugaku). We build the HPL benchmark and Stream benchmark following the same approach for the \mcimone~cluster on two SoA computing nodes, namely the Marconi100\cite{m100}(ppc64le, IBM Power9) computing node and the and Armida\cite{armida} computing node (ARMv8a, Marvell ThunderX2) and compared the attained FPU utilization as a metric of efficiency against the one obtained by \mcimone~while keeping the same benchmarking boundary conditions (e.g.: vanilla, unoptimized libraries and software stack deployed via a popular package manager). 
    Results show that upstream HPL achieved $46.5\%$ utilization on \mcimone, the Marconi100\cite{m100} and Armida\cite{armida} compute nodes achieved $59.7\%$ and $65.79\%$ of their peak respectively. The \mcimone~node achieves slightly lower FPU utilization but in the range with the state of the art. When running an unoptimized Stream benchmark, \mcimone~obtained just the $15.5\%$ of the peak bandwidth, while Marconi100 and Armida obtained an efficiency of $48.2\%$ and $63.21\%$ respectively, pointing to significant margins for improvement in application and software stack tuning to the hardware target.

    \item We extended the ExaMon monitoring framework \cite{bartolini_paving_2019} to monitor the \mcimone~ cluster. We characterised the power consumption of various applications executed on \mcimone. We reported a power consumption of $4.81W$ in idle, composed of $64\%$ of core power, $13\%$ related to DDR and $23\%$ of related to PCI subsystem. During CPU intensive benchmark run the SiFive Freedom U740 SoC we reported a power consumption of $5.935W$, composed of $69\%$ of core power, $14\%$ related to DDR and $18\%$ related to PCI subsystem. By profiling the power consumption of the core complex during the boot process we measured a $0.981W$ of leakage only power ($32\%$ of the Idle power) and measured $0.514W$ of power consumed by the operating system during idle ($17\%$ of the Idle power) and a remaining $1.577W$ of dynamic and clock tree power, accounting for the $51\%$ of the core idle power. In addition to providing a detailed analysis of power consumption, ExaMon enabled us to detect and mitigate a thermal design issue in the cluster. 
    
\end{itemize}

\section{Related Works}  \label{sec:rw}

The most recent successful effort to introduce a new ISA to HPC has involved the ARM ISA. Bringing the Arm ISA and software ecosystem to HPC maturity has required almost a decade and several funding rounds: The Mont-Blanc EU project series started in 2011, leading to the first ARM-based HPC cluster deployed in 2015\cite{rajovic_mont-blanc_2016}, based on SoCs developed for the embedded computing market. Notably, since June 2020,  Fugaku\cite{sato_co-design_2020}, the fastest supercomputer in TOP500, is based on ARM scalable vector extension (SVE) ISA, and achieves more than 400 PFLOPs. Further, high-performance ARM-based SIMD processors are being adopted in servers and datacenters worldwide. We observe that it took approximately a decade for ARM to become a strong player in these highly competitive markets, even though  X86 is still by far the dominant architecture in HPC and cloud.

The RISC-V ISA has been conceived just a decade ago, thus clearly its market penetration is much smaller than the incumbent ARM and X86 ISAs. Today, only a few 64-bit RISC-V (RV64G ISA) SoCs are available commercially and none is in volume production for HPC or performance servers. Nevertheless, several high-performance RISC-V processors have been announced for high-performance general-purpose and accelerated computing markets 
\cite{RIVOSIC,SEMIDYNAMICS,ESPERANTO}. In addition, a few research prototypes have been presented in the recent literature that demonstrate on silicon the technical feasibility and competitiveness of high-performance RISC-V computing engines \cite{zaruba_manticore_2020,chen2021,Schmidt2022}. Furthermore, the European Processor Initiative (EPI) launched in 2019 is funding a major research thrust to develop RISC-V based accelerators for HPC \cite{EPI}.

Among the RV64G chips available in low volumes on the market, for our cluster we chose the SiFive Freedom U740 SoC, featuring a 64-bit dual-issue, superscalar RISC-V U7 core complex configured with  four U74 cores and one S7 core, an  integrated high speed DDR4 memory controller, a root complex PCI Express Gen 3 x8 and standard peripherals. The availability of a main memory interface with reasonable performance and a PCIe root complex for connecting fast storage, IOs and accelerators, makes this SoC a good basis for exploring the deployment of RISC-V processors in a scalable cluster and working on the software stack. Still, it is apparent that the performance and number of cores in the SoC is not sufficient to achieve performance comparable to mature ARM and X86 cores.

The maturity of the software ecosystem around RISC-V has been growing at a very fast rate. A reasonably complete snapshot of major software packages available for RISC-V is maintained by the RISC-V foundation \cite{riscvxchange}.
While the list is not complete, due to the very fast growth of the RISC-V community of developers, it is clear that porting efforts so far have mainly focused on embedded and AI applications. A HPC special interest group (SIG) for RISC-V has been founded in 2019 \cite{hpcsig}. However, to the best of our knowledge, the demonstration of a complete software stack and HPC applications running on real hardware on RISC-V nodes in a multi-blade cluster is still missing. The present work aims at filling this gap. 

In addition to libraries and tools for HPC application deployment, a pro\-du\-ction-\-ready HPC system must support fine-grain utilization, performance and power monitoring of the computing resources to enable efficient computing, power, thermal management and anomaly detection for reliability. Recently, several works have been proposed to extend the power monitoring attainable from the voltage regulator modules leveraging shunt resistors, current probes, and out-of-band telemetry\cite{libri_paella_2020}.  In addition, Operational Data Analytics\cite{netti_conceptual_2021} (ODA) has been introduced focusing on  monitoring and managing large scale HPC installations. In this area, vertical solutions encompassing all layers (from data gathering and storage to processing and analysis) have been proposed.
Bautista et al. \cite{bautista2019collecting} describe an infrastructure for extreme-scale operational data collection, known as OMNI. In \cite{bartolini_paving_2019} Bartolini et al. describe ExaMon, an ODA infrastructure leveraging: i) Distributed sensing plugins (including node-level metrics, processing elements performance metrics, dedicated fine-grain power monitoring meters, facility data); ii) Scalable storage backends; iii)  Visualization and analytics targetting anomaly detection and intrusion detection systems. Current ODA tools are available only for the dominant ARM and X86 environments. In this work we advance the stat of the art demonstaring a fully-operational port of the Examon ODA infrastructure to the \mcimone{} RISC-V cluster. 

\section{\mcimone~Hardware}  \label{sec:hw}
\begin{figure*}
    \begin{minipage}{.4\linewidth}
    \centering
    \includegraphics[width=\linewidth]{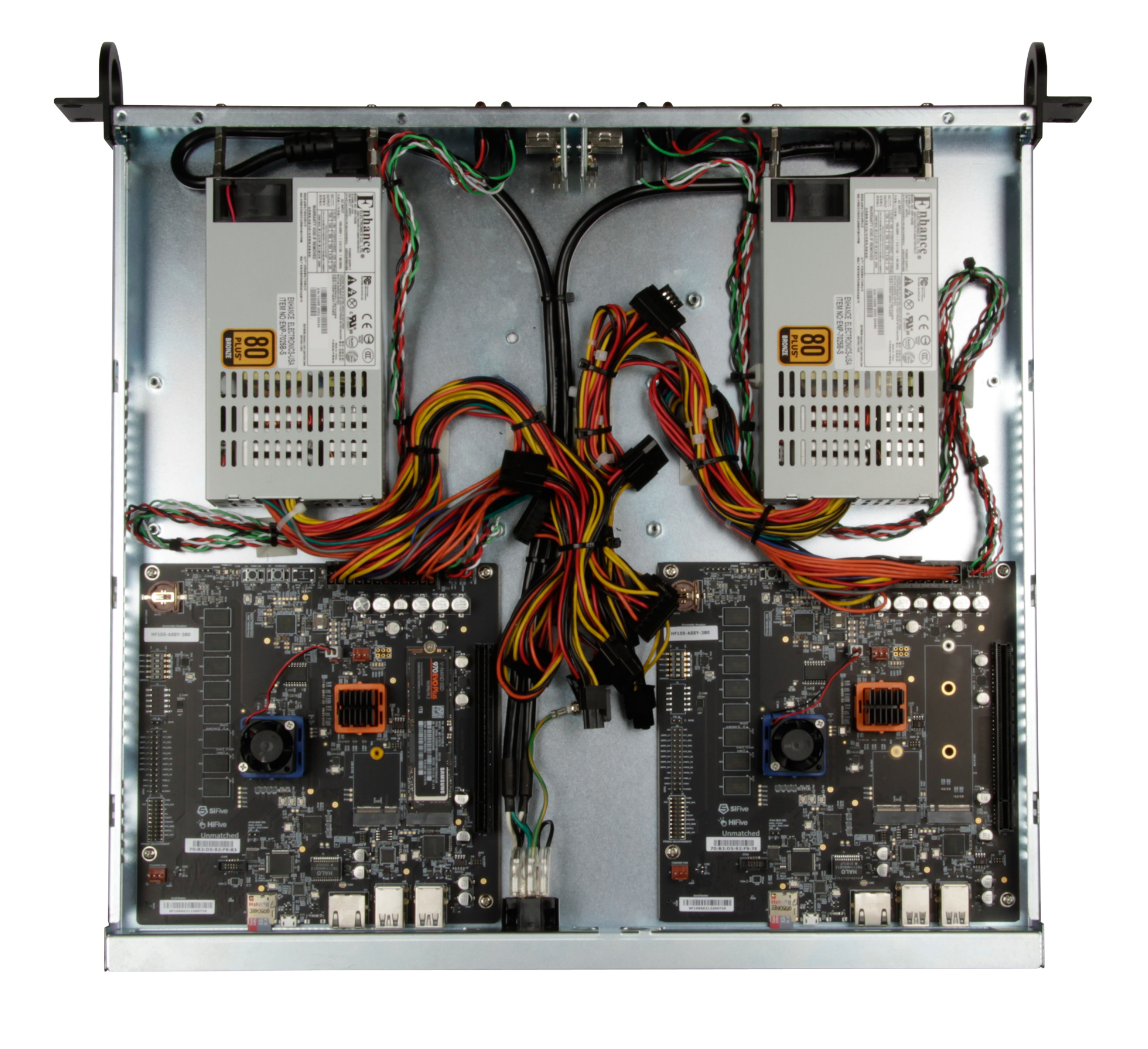}
    \end{minipage}
    \begin{minipage}{.6\linewidth}
    \centering
    \includegraphics[width=\linewidth]{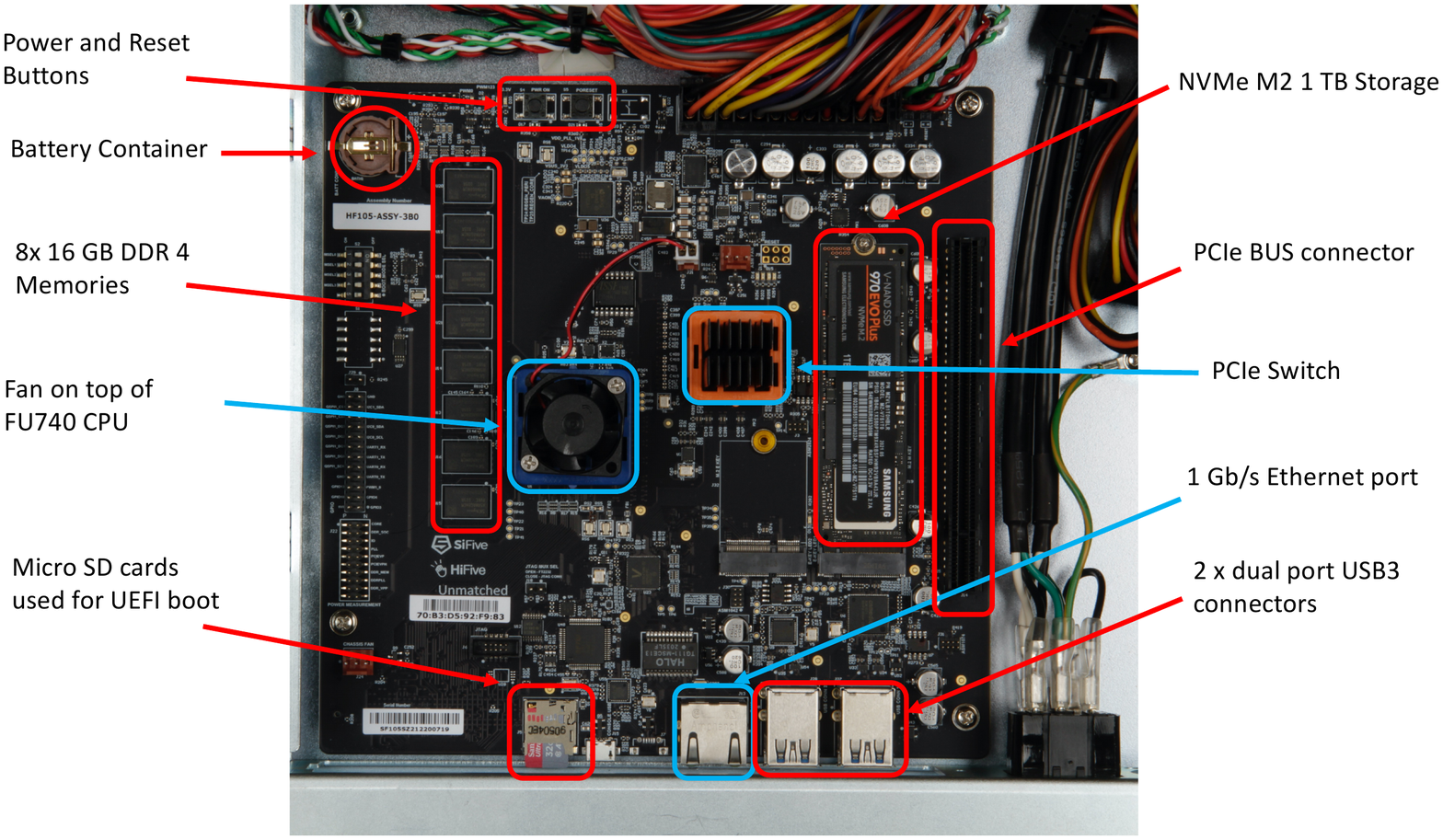}
\end{minipage}
\caption{The E4 RV007 Server Blade is based on a dual SiFive Freedom U740 SoC, the form factor is \SI{4.44}{\centi\meter} (1 RackUnit)  high, \SI{42.5}{\centi\meter} width,  \SI{40}{\centi\meter} deep. The size of each RISC-V development board is \SI{170}{\milli\meter} per \SI{170}{\milli\meter}.}
\label{fig:serverDetail}
\end{figure*}

\mcimone{} is based on the SiFive Freedom U740 RISC-V SoC  HiFive Unmatched board integrated in an HPC node form factor. The E4 RV007 blade prototype system, adopted as \mcimone~building Block, is a dual-board platform server, with a form factor of \SI{4.44}{\centi\meter} (1 RackUnit) high, \SI{42.5}{\centi\meter} width, \SI{40}{\centi\meter} deep. Two \SI{250}{\watt} power supplies, one for each board (compute node), are installed inside the case. This allows to turn on every single compute node individually, see Figure \ref{fig:serverDetail}, and makes the system ready with abundant power headroom for future expansions with hardware accelerators and PCIe Network Card connector. 

The board follows the Industry Standard Mini-ITX with a size of \SI{170}{\milli\meter} per \SI{170}{\milli\meter}. 
It features one SiFive Freedom U740 SoC, 16 GB of 64-bit DDR4 memory operating up to 1866s MT/s and high-speed interconnects with PCIe Gen 3 x16 (but it's limited to x8 lanes), one Gigabit Ethernet, and four USB 3.2 Gen 1, see Figure \ref{fig:serverDetail}.

In RV007 node the M.2 M-key expansion slot is occupied by a \SI{1}{\tera\byte} NVME 2280 SSD Module storage device used by the Operating System. The Micro SD card is present and used for the UEFI Boot.
Two buttons for reset and power up operations are available on top of the board and in front of the case.


The FU740-C000 is a Linux-capable SoC powered by SiFive’s U74-MC, the first (to the best of our knowledge) commercially available superscalar heterogeneous multi-core RISC-V Core Complex. The FU740-C000 is compatible with all applicable RISC‑V standards.

The U74-MC core complex is composed of four 64-bit U74 RISC‑V (Application) cores. Each U74 core has a dual issue in-order execution pipeline, with a peak sustainable execution rate of two instructions per clock cycle. The U74 core supports Machine, Supervisor, and User privilege modes as well as standard Multiply, Single-Precision Floating Point, Double-Precision Floating Point, Atomic, and Compressed RISC‑V extensions.

The SiFive Freedom board features a Microsemi VSC8541 chip to interconnect the SiFive Freedom U740 SoC with a single port gigabit Ethernet copper interface. Moreover, we equipped two of the compute nodes with an Infiniband Host Channel Adapter (HCA) widely used in large-scale HPC systems. We target an Infiniband FDR HCA (56Gbit/s) to leverage RDMA communications among different nodes to improve the network throughput and the communication latency. We used a Mellanox ConnectX-4 FDR HCA interconnect through the PCI-e interface available on the compute node. This HCA support x8 free PCIe Gen 3 lanes, which are currently supported by the vendor. The first experimental results show that the kernel is able to recognise the device driver and mount the kernel module to manage the Mellanox OFED stack. We are not able to use all the RDMA capabilities of the HCA due yet-to-be-pinpointed incompatibilities of the software stack and the kernel driver. Nevertheless we successfully run an IB \texttt{ping} test between two boards and between a board and an HPC server showing that full Infiniband support could be feasible. This is currently a feature under development.


In addition, the SiFive Freedom U740 SoC features 7 separated power rails including the core complex, IOs, PLLs, DDR subsystem and PCIe one. The HiFive Unmatched board implements separated shunt resistors in series with each of the SiFive U740 power rails as well as for the on-boards memory banks \cite{sifive_2021}.  



\section{\mcimone~Software Stack}  \label{sec:sw}

Since our goal was to build a software environment as close as possible to a production HPC cluster, we leveraged the Spack \cite{gamblin_spack_2015} package manager to deploy the full software stack and make it available to all system users via environment modules \cite{Furlani1991ModulesP}. Actual Spack architecture and micro-architecture support, in the form of platform-specific toolchain flags, is provided by the \texttt{archspec} \cite{culpo_archspec_2020} module. Explicit support for the \texttt{linux-sifive-u74mc} target triple was already present (\texttt{archspec} version 0.1.3) and tested to be working without modifications. The user-facing software stack installed successfully via Spack (version 0.17.0) and presented to users is listed in Table~\ref{table:swtack} (transitive dependencies omitted for brevity). All of the nodes are running upstream Ubuntu 21.04 deployed from \texttt{riscv64} server images without modifications and mount a remote NFS.

\begin{table}
    \centering
    \caption{User-facing software stack deployed on \mcimone}
    \label{table:swtack}
    \vspace{0.1in}
        \begin{tabular}{@{}lr@{}}
        \toprule
        \multicolumn{1}{l}{Package} & \multicolumn{1}{l}{Version} \\
        \midrule
        \texttt{gcc}              & 10.3.0 \\
        \texttt{openmpi}          & 4.1.1 \\
        \texttt{openblas}         & 0.3.18 \\
        \texttt{fftw}             & 3.3.10 \\
        \texttt{netlib-lapack}    & 3.9.1 \\
        \texttt{netlib-scalapack} & 2.1.0 \\
        \texttt{hpl}              & 2.3 \\
        \texttt{stream}           & 5.10 \\
        \texttt{quantumESPRESSO}  & 6.8 \\
        \bottomrule
    \end{tabular}

\end{table}

\subsection{System Software}
We ported on \mcimone~ all the essential services needed for running HPC workloads in a production environment, namely NFS, LDAP and the SLURM job scheduler. Porting all the necessary software packages to RISC-V was relatively straightforward, and we can hence claim that there is no obstacle in exposing \mcimone~ as a computing resource in a HPC facility. However, full integration requires integrating \mcimone~ within  a holistic monitoring framework. For that purpose we use the ExaMon framework\cite{bartolini_paving_2019}. In the next sub-section we describe its configuration and the measured metrics.

\subsection{ExaMon Configuration}\label{AA}
The typical configuration of ExaMon consists in installing plugins dedicated to data sampling, a broker for transport layer management and a database for storage. For \mcimone~ cluster both broker and database are installed in their basic configuration on a master node, while plugins have been specifically developed/adapted for this project and installed on the compute nodes.
As a first step, we created a dedicated version of the \textit{pmu\_pub} \cite{examon_date} plugin to acquire the performance counters available in the Linux OS through the \textit{perf\_events} interface. In the current version of the Kernel the RISC-V architecture provides, through this interface, the fixed INSTRET and CYCLE counters. By default, the remaining programmable counters available on the hardware performance monitoring (HPM) unit of the U740 SoC \cite{sifive_2021} are disabled at boot time. We have therefore developed a patch for the bootloader (U-Boot) useful to enable and program all counters.

The counters are sampled for each core of the SoC in user-mode by the \textit{pmu\_pub} plugin at regular intervals (2Hz) and the values are sent to the MQTT transport layer. The data model adopted for this application follows the ExaMon specification and consists in the definition of the MQTT topic and payload as described in the Table \ref{tab:examon_formats}. 

A second plugin has been installed and configured to collect operating system statistics, \textit{stats\_pub}. This plugin mainly accesses the \textit{sysfs} and \textit{procfs} filesystems to get useful metrics about system resources such as load, CPU usage, memory usage, network bandwidth and other metrics as described in Table \ref{tab:stats_metrics}. In particular, the HiFive Unmatched board is equipped with three thermal sensors dedicated respectively to the SoC, the Motherboard and the NVME SSD. These sensors are available through the \textit{hwmon sysfs} interface as shown in Table \ref{tab:sysfs}. This plugin samples data with a frequency of 0.2Hz.

Finally, the data collected for each board are available to be viewed and processed through the various interfaces provided by ExaMon. Through an instance of Grafana\cite{bartolini_paving_2019} connected to the database it is possible to visualize the trend of the metrics in real time, during the execution of the benchmark. The data can also be analyzed in batch mode using scripts and accessing the database through the dedicated RESTful API over HTTP.  

\begin{table}[]
\scriptsize
\centering
\caption{ExaMon: Topic and payload formats}
\label{tab:examon_formats}
\begin{tabular}{lll}
\hline
Plugin & Topic & Payload \\ \hline
pmu\_pub & \begin{tabular}[c]{@{}l@{}}org/XXXXXX/cluster/XXXXXXX/\\ node/\textless{}hostname\textgreater{}/plugin/\\ pmu\_pub/chnl/data/core/\\ \textless{}id\textgreater{}/\textless{}metric\_name\textgreater{}\end{tabular} & \textless{}value\textgreater{};\textless{}timestamp\textgreater{} \\ \hline
stats\_pub & \begin{tabular}[c]{@{}l@{}}org/XXXXXX/cluster/XXXXXXX/\\ node/\textless{}hostname\textgreater{}/plugin/\\ dstat\_pub/chnl/data/\\ \textless{}metric\_name\textgreater{}\end{tabular} & \textless{}value\textgreater{};\textless{}timestamp\textgreater{} \\ \hline
\end{tabular}
\end{table}


\begin{table}[]
\centering
\caption{Metrics collected by the stats\_pub plugin}
\label{tab:stats_metrics}
\begin{tabular}{ll}
\hline
Type & Metric \\ \hline
Load & load\_avg.1m,load\_avg.5m,load\_avg.15m \\
I/O & io\_total.read,io\_total.writ \\
Processes & procs.run,procs.blk,procs.new \\
\multirow{3}{*}{Memory} & memory\_usage.used,memory\_usage.free,memory\_usage.buff, \\
 & memory\_usage.cach \\
 & paging.in,paging.out \\
Disk & dsk\_total.read,dsk\_total.writ \\
System & system.int,system.csw \\
\multirow{2}{*}{CPU} & total\_cpu\_usage.usr,total\_cpu\_usage.sys,total\_cpu\_usage.idl, \\
 & total\_cpu\_usage.wai,total\_cpu\_usage.stl \\
Network & net\_total.recv,net\_total.send \\
\multirow{2}{*}{Temperatures} & temperature.mb\_temp,temperature.cpu\_temp, \\
 & temperature.nvme\_temp \\ \hline
\end{tabular}
\end{table}

\begin{table}[]
\centering
\caption{Sysfs entries for the temperature sensors}
\label{tab:sysfs}
\begin{tabular}{ll}
\hline
Sensor & Sysfs Files \\ \hline
nvme\_temp & /sys/class/hwmon/hwmon0/temp1\_input \\
mb\_temp & /sys/class/hwmon/hwmon1/temp1\_input \\
cpu\_temp & /sys/class/hwmon/hwmon1/temp2\_input \\ \hline
\end{tabular}
\end{table}

\section{Experimental Results}  \label{sec:exp}
In this section, we report the characterisation of the \mcimone~ Cluster and of its software stack with the objective of assessing its maturity.
In subsection \ref{subsec:app_perf} we focus on the software stack by compiling and running three different applications without manual optimisations.
This lets us assess the available toolchains and libraries' capability to extract the application's performance given the new in HPC RISC-V ISA.
We will then focus in subsection \ref{subsec:power_characterization} one the power characterisation of one compute node.
Finally, in subsection \ref{subsec:examon_dasboard} we will describe cluster-level performance based on live dashboards extracted by ExaMon.

\subsection{Application performance}
\label{subsec:app_perf}

\begin{figure}
        \centering
        \includegraphics[width=1\linewidth]{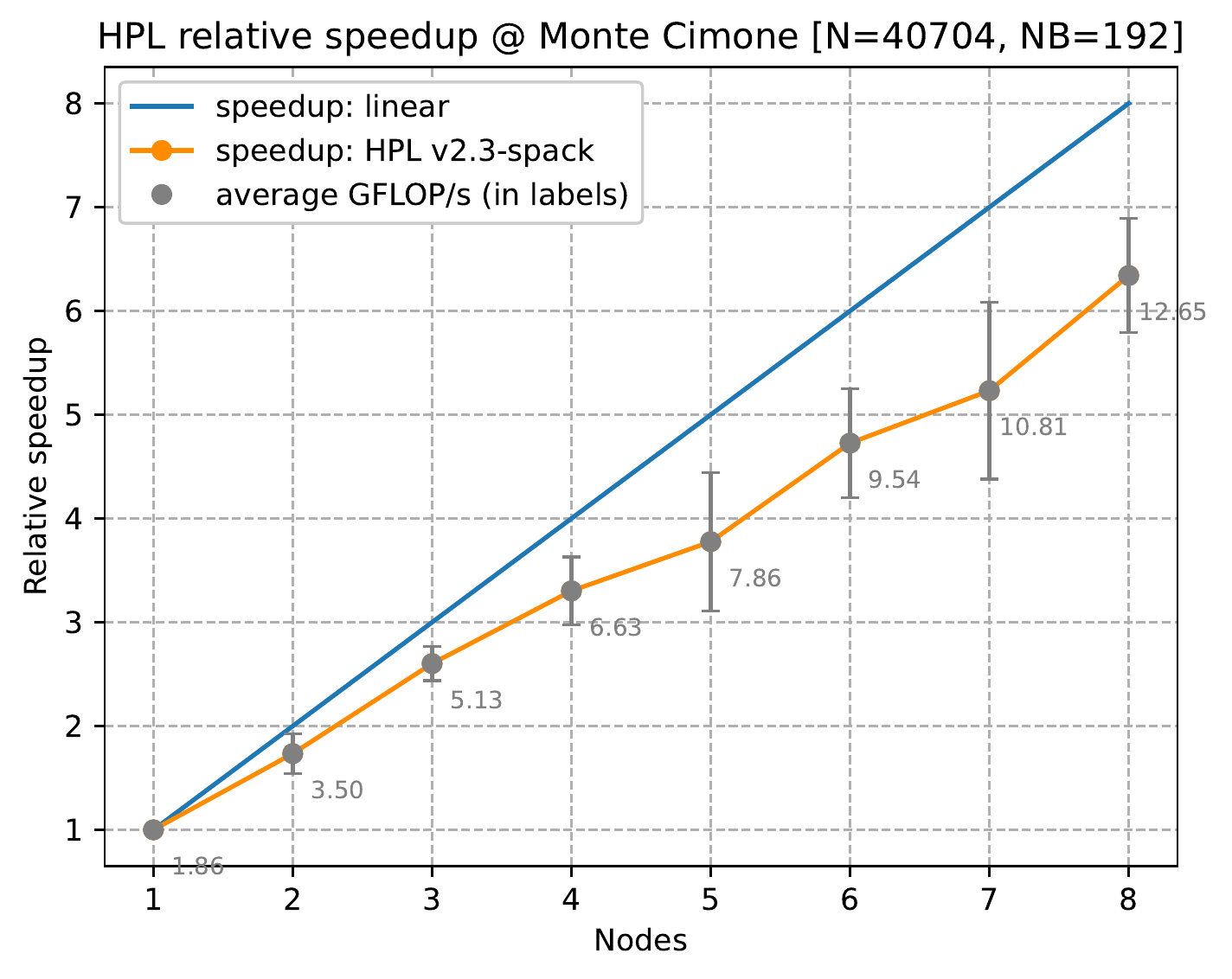}
        \caption{HPL strong scaling tests on \mcimone. Average attained throughput values are shown in labels. Standard deviations are calculated on 10 repetitions.}
        \label{fig:hpl_strong_scaling}
\end{figure}

Considering the peak theoretical value of 1.0 GFLOP/s/core, inferred from the micro-architecture specification\cite{sifive_2021}, leading to a 4.0 GFLOP/s peak value for a single chip, the upstream HPL \cite{petitet2008} benchmark (built on top of the software stack shown in Section~\ref{sec:sw}) reached a sustained value of \(1.86 \pm 0.04\) GFLOP/s on a single node (on a \texttt{N=40704} and \texttt{NB=192} HPL configuration and a total runtime of \(24105 \pm 587\) s); this amounts to 46.5\% of the theoretical peak, a result we deem to be promising considering the upstream, unmodified software stack used in this phase. The same experiment, run on both the Marconi100\cite{m100} system at Cineca and the Armida\cite{armida} system at E4 using the same upstream software stack (and no vendor libraries) with the same MPI topology of 1 MPI task per physical core attained 59.7\% and 65.79\% of a single node's CPU-only theoretical peak respectively, a result that is comparable to what we observed on \mcimone. The same HPL configuration has been used to carry out a \mcimone full-machine benchmark experiment leveraging the 1 Gb/s network currently available, reaching a sustained value of \(12.65 \pm 0.52\) GFLOP/s using all of the eight nodes (with a total runtime of \(3548 \pm 136\) s); this amounts to 39.5\% of the entire machine's theoretical peak and to 85\% of the extrapolated attainable peak in case of perfect linear scaling from the single-node case. Relative speedup obtained during the HPL strong scaling experiment are shown in Figure~\ref{fig:hpl_strong_scaling}. Again, we consider these results to be promising and deserving both further optimization on the software side and tuning (or technology upgrade) on the interconnect side.

The STREAM \cite{McCalpin1995} benchmark has been used to measure the attainable memory bandwidth on a single node. Out of the peak 7760 MB/s \cite{sifive_2021}, a 4-thread experiment measured the values shown in Table~\ref{table:stream}. Being the node a UMA system, no topology configuration had to be taken into account. We consider the results attained via upstream, unmodified STREAM unsatisfactory: the results on \mcimone show an attained bandwidth of no more than 15.5\% of the available peak bandwidth. The same experiment involving an upstream, unoptimized STREAM benchmark ran on both Marconi100\cite{m100} and Armida\cite{armida} (using the same topology with 1 OpenMP thread per physical core) attained 48.2\% and 63.21\% of the peak bandwith respectively, suggesting that a result higher than the lower quartile should be easily attained with little to no effort. This observation is worth of further experimentation, in particular:
\\(i) the L2 prefetcher provided by the micro-architecture\cite{sifive_2021}, being able of tracking up to eight streams per core, should be perfectly capable of reducing the gap between the two experiments shown in Table~\ref{table:stream} (DDR-bound and L2-bound) given the large degree of spatial and temporal locality shown by the STREAM memory access patterns. Further analysis is needed to understand how the prefetcher is currently operating and the modifications needed to leverage it properly;
(ii) the overall data size used by STREAM is currently limited by the RISC-V code model. The \texttt{medany} code model used by \texttt{RV64} requires that every linked symbol resides within a \(\pm 2 GiB\) range from the \texttt{pc} register \cite{sifive_2021,risc-v_international_risc-v_2021}. Since the upstream, unmodified STREAM benchmark uses statically-sized data arrays in a single translation unit preventing the linker to perform \textit{relaxed} relocations, their overall size cannot exceed 2 GiB. Further experiments on available workarounds for the absence of a \textit{large code model} \cite{sifive_risc-v_nodate} and modifications to the STREAM source itself to overcome this limitation are needed;
\\(iii) while the architecture provides both the \texttt{Zba} and \texttt{Zbb} RISC-V bit manipulation standard extensions\cite{sifive_2021}, the upstream GCC 10.3.0 toolchain isn't capable of emitting them nor the underlying GNU \texttt{as} assembler (shipped with GNU Binutils 2.36.1) is able to properly assemble them. Experiments with the latest upstream GCC version (\textit{minimal} support for bit manipulations code generation landed in GCC 12 \cite{gcc_commit_bitmanip_support}) and the upstream development version of GNU Binutils (patches already merged\cite{binutils_commit_bitmanip_support}, expected to be shipped with GNU Binutils 2.37.x) are needed to assess its impact on current STREAM measurements.

\begin{table}
    \centering
    \caption{STREAM, 4 threads} 
    \label{table:stream}
    \vspace{0.1in}
    
    \begin{tabular}{@{}lrr@{}} 
    \toprule
        \multirow{2}{*}{Test} &
        \multicolumn{1}{c}{\texttt{STREAM.DDR}} &
        \multicolumn{1}{c}{\texttt{STREAM.L2}} \\

        \cmidrule(lr){2-3}
        &
        1945.5 MiB [MB/s] &  
        1.1 MiB [MB/s] \\  
        
        \midrule
        
        \texttt{copy}  & $1206 \pm 3.26$ & $7079 \pm 2.11$ \\
        \texttt{scale} & $1025 \pm 4.94$ & $3558 \pm 3.72$ \\
        \texttt{add}   & $1124 \pm 4.93$ & $4380 \pm 3.72$ \\
        \texttt{triad} & $1122 \pm 5.63$ & $4365 \pm 3.56$ \\

        \bottomrule
    \end{tabular}
    
\end{table}

Regarding user applications, we carried out benchmarks for the quantum\-ES\-PRES\-SO \cite{QE} suite, in particular using its LAX test driver, compiled with OpenMPI, that performs a blocked (and optionally distributed) matrix diagonalization as a benchmark representative of the full-scale application workload. For a \(512^2\) input matrix we obtained a value of \(1.44 \pm 0.05\) GFLOP/s (36\% of the theoretical FPU efficiency) on a single node over a total test duration of \(37.40 \pm 0.14\) s.


\subsection{Power characterization}
\label{subsec:power_characterization}
We characterised the system's power consumption under test, exploiting the set of nine power lines available on-boards with embedded shunt resistors for current monitoring.

\begin{figure}
    \centering
    \includegraphics[width=\linewidth]{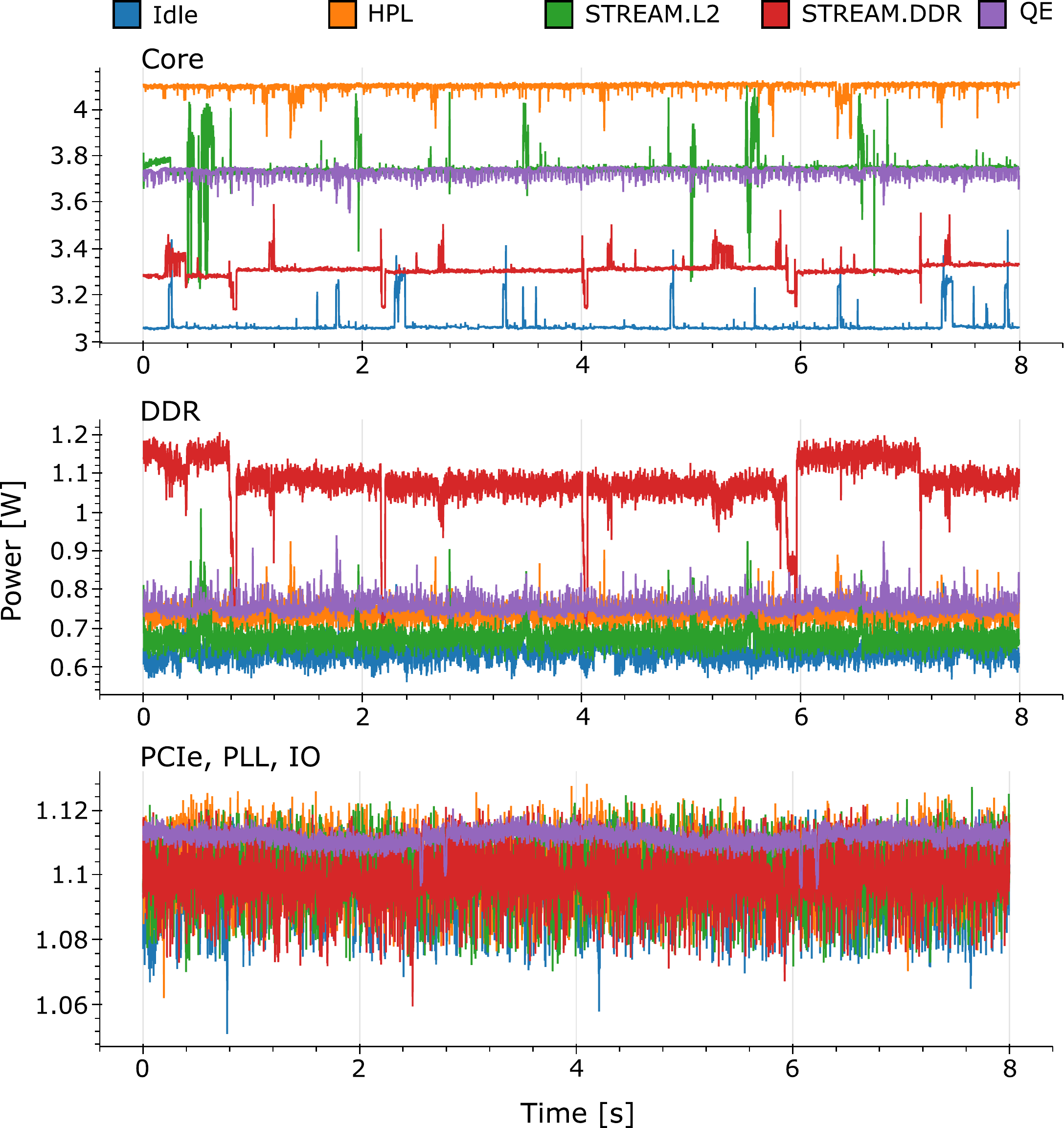}
    \caption{
    Snapshot of the power consumption of the core (top), the DDR (middle) and the PCIe, PLL and IO subsystems (bottom).
    The traces are obtained observing power consumption for 8 seconds during benchmark run and averaging raw data using 1 ms windows.
    }
    \label{fig:power_traces}
\end{figure}

Power consumption of a cluster node is characterised using a set of standard HPC benchmarks run on a single node with the maximum allowed parallelism.
Additionally, we measured the system's power consumption in idle, when only normal OS services and daemons are running in the background to evaluate the impact of benchmark running on power consumption.
Power measurement results are collected in Table \ref{table:power_consumption}. Figure \ref{fig:power_traces} reports 8 seconds of power traces for each of the benchmark executed.

\begin{table*}
    \centering
    \caption{Power consumption} 
    \label{table:power_consumption}
    \vspace{0.1in}
    
    \begin{tabular}{@{}lrrrrrrrrrrrr@{}} 
    \toprule
        \multirow{3}{*}{Line} &
        \multicolumn{2}{c}{\multirow{2}{*}{\texttt{Idle}}} &
        \multicolumn{2}{c}{\multirow{2}{*}{\texttt{HPL}}} &
        \multicolumn{2}{c}{\multirow{2}{*}{\texttt{STREAM.L2}}} &
        \multicolumn{2}{c}{\multirow{2}{*}{\texttt{STREAM.DDR}}} &
        \multicolumn{2}{c}{\multirow{2}{*}{\texttt{QE}}} &
        \multicolumn{2}{c}{\texttt{Boot}} \\
        
        &
        & &                                                                   
        & &                                                                   
        & &                                                                   
        & &                                                                   
        & &                                                                   
        \multicolumn{1}{c}{\texttt{R1}} & \multicolumn{1}{c}{\texttt{R2}} \\  
        
        \cmidrule(lr){2-3}
        \cmidrule(lr){4-5}
        \cmidrule(lr){6-7}
        \cmidrule(lr){8-9}
        \cmidrule(lr){10-11}
        \cmidrule(lr){12-13}
        
        & 
        [mW] & [\%] &   
        [mW] & [\%] &   
        [mW] & [\%] &   
        [mW] & [\%] &   
        [mW] & [\%] &   
        [mW] & [mW] \\  
        
        \midrule
        
        \texttt{core}      & 3075 &  64 & 4097 &  69 & 3714 &  68 & 3287 &  62 & 3825 &  67 &  984 & 2561 \\
        \texttt{ddr\_soc}  &  139 &   3 &  177 &   3 &  170 &   3 &  232 &   4 &  176 &   3 &   59 &  197 \\
        \texttt{io}        &   20 &   0 &   20 &   0 &   20 &   0 &   20 &   0 &   20 &   0 &    5 &   20 \\
        \texttt{pll}       &    1 &   0 &    1 &   0 &    1 &   0 &    1 &   0 &    1 &   0 &    0 &    2 \\
        \texttt{pcievp}    &  521 &  11 &  527 &   9 &  524 &  10 &  522 &  10 &  530 &   9 &   12 &  231 \\
        \texttt{pcievph}   &  555 &  12 &  554 &   9 &  554 &  10 &  555 &  10 &  561 &  10 &    1 &  395 \\
        \texttt{ddr\_mem}  &  404 &   8 &  440 &   7 &  401 &   7 &  592 &  11 &  434 &   8 &  275 &  467 \\
        \texttt{ddr\_pll}  &   28 &   1 &   28 &   1 &   28 &   1 &   28 &   1 &   28 &   1 &    0 &   29 \\
        \texttt{ddr\_vpp}  &   67 &   1 &   90 &   2 &   73 &   1 &   98 &   2 &   95 &   2 &   49 &  122 \\
        \cmidrule(l){2-13} 
        Total              & 4810 & 100 & 5935 & 100 & 5486 & 100 & 5336 & 100 & 5670 & 100 & 1385 & 4024 \\
        
        \bottomrule
    \end{tabular}
    
\end{table*}

The power required by the system to run is comprised between 4.810 Watts, in idle, and 5.935 Watts when the most power-hungry computation is run.
Most of the system consumption is due to the core subsystem, which absorbs 3.543 Watts on average, reaching a peak consumption of 4.097 Watts for CPU intensive benchmarks such as HPL.
The results show two more main sources of power consumption.
i) The PCIe subsystem consistently requires 1 Watt, roughly 20\% of system consumption, even if nothing is attached to the HiFive Unmatched PCIe connector.
ii) DDR4 memory requires between 0.638 Watts when the system is idle and 0.950 Watts when the STREAM benchmark is run with a data size sufficient to disrupt L2 data locality.
In general, DDR memory subsystem power consumption sits between 12\% and 18\% of the overall.
The PLL subsystem and the IO interfaces together stand below 1\% of the overall consumption for the tested workloads. 

\begin{figure}
    \centering
    \includegraphics[width=\linewidth]{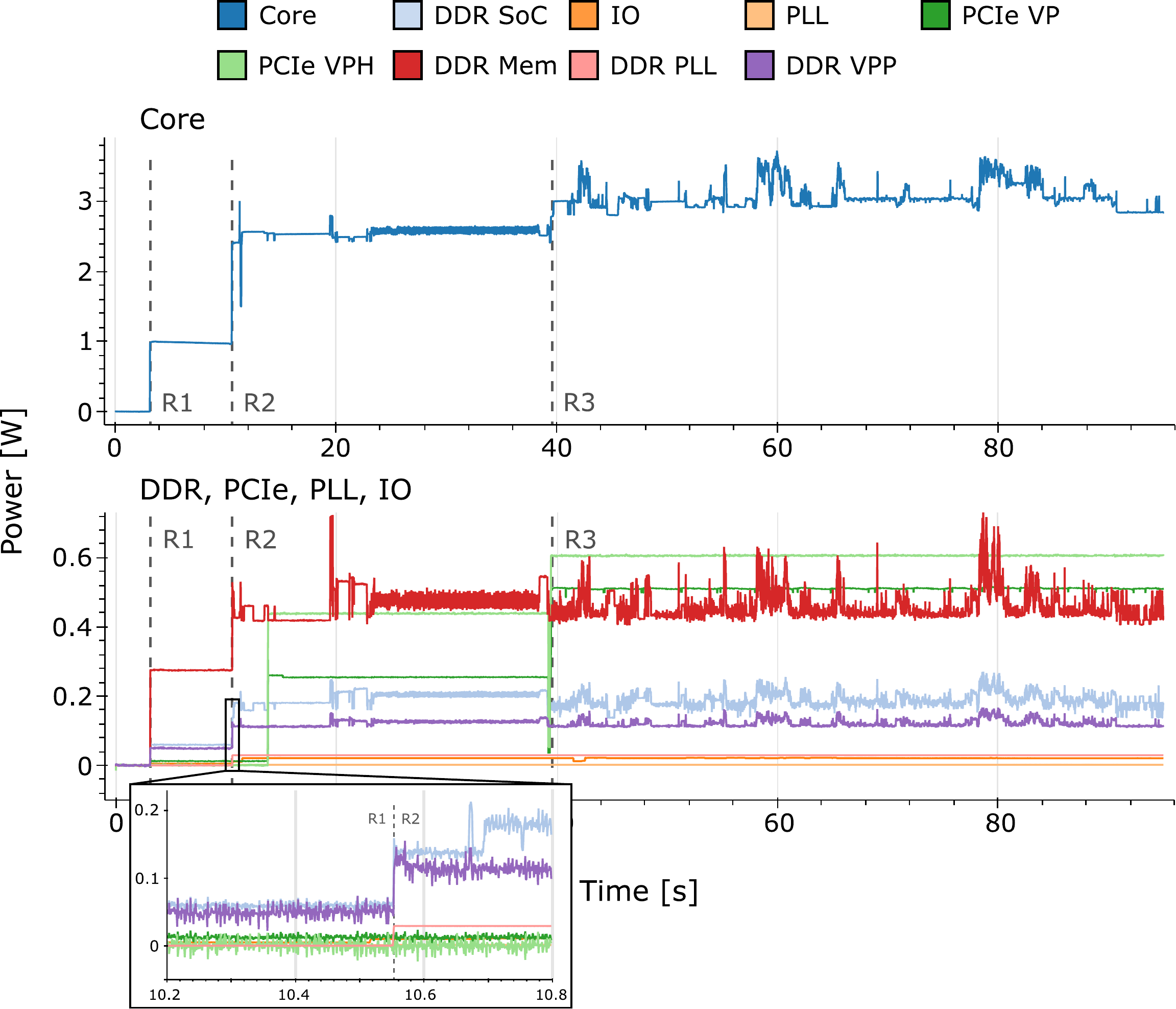}
    \caption{
    Power consumption for the Core (top), DDR, PCIe, PLL and IO (bottom) subsystems during system boot.
    Boot phases: power-on (R1), bootloader (R2), O.S. boot (R3).
    The Figure also shows the detail of PLL activation.
    }
    \label{fig:boot}
\end{figure}

Figure \ref{fig:boot} reports 80 seconds of power traces measured during the boot process. It is interesting to note a region of power consumption ($4s<t<10s$) at which the core complex it is powered on, but PLL (reported in yellow) is not active yet, we call these regions, \textit{R1}. The average power consumption of the core complex in that region is 0.984 Watt.

As soon as the PLL activates, the power consumption jumps to a value of 2.561 Watts (\textit{R2}) which increases to the value of 3.082 Watts, comparable with idle power for t $> 40s$ \textit{R3}. These three regions are of interest as they allow us to estimate the three components of the power consumption, which are hard to extract from a commercial off-the-shelf device without complex laboratory equipment. As in region R1, only the power supply but no clock is applied to the core complex, which is consuming only leakage power, which accounts for 32\% of the idle power. In region R2, the clock is propagated to the core complex, but the operating system is not yet loaded, memory is initialising, and boot-loader tasks are ongoing. This power consumption accounts mainly for the clock tree and core's dynamic power. In region R3, the operating system is executing, but no active workload is in execution. 

We can thus conclude that the operating system power accounts for the gap between R3 and Idle power (3.072 Watts) and R2 power consumption (2.561 Watts), which is (0.514 Watts) the 17\% of the idle power. Conversely the difference between R2 and R1 accounts for the dynamic
and clock tree power, which is 1.577 Watts equal to the 51\% of the core idle power. By focusing to the DDR subsystem (ddr\_mem) we can make the similar consideration having in R1 0.275 Watts of leakage power, which is the 68\% of their idle power the remaining part 32\% is expected to be self-refresh and O.S. accesses for house keeping during O.S. idle period.

\subsection{Temperature monitoring}
\label{subsec:examon_dasboard}

\begin{figure}
    \centering
    \includegraphics[width=\linewidth]{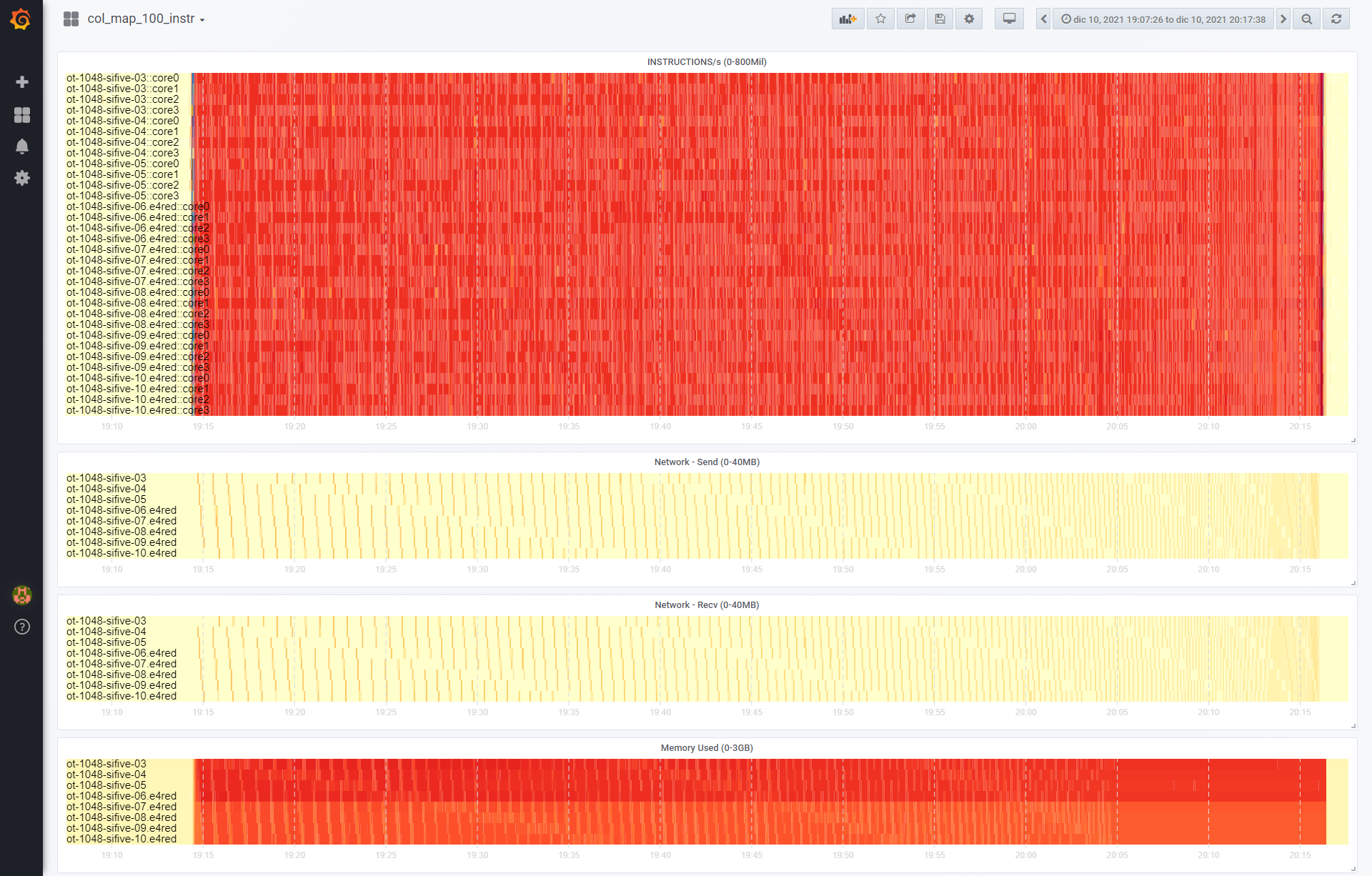}
    \caption{ExaMon - HPL Heatmap: Instructions/s, Network traffic, Memory Usage}
    \label{fig:HPL}
\end{figure}

We used the ExaMon monitoring subsystem to observe and report the cluster's activity to pinpoint inefficiencies and  find opportunities for performance, power and thermal optimization.
Figure~\ref{fig:HPL} reports the nodes activity during the HPL run. From it we can identify the communication patterns, corresponding to a lower instruction count. We can expect to achieve higher performance once the RDMA will be supported over infiniband.

\begin{figure}
    \centering
    \includegraphics[width=\linewidth]{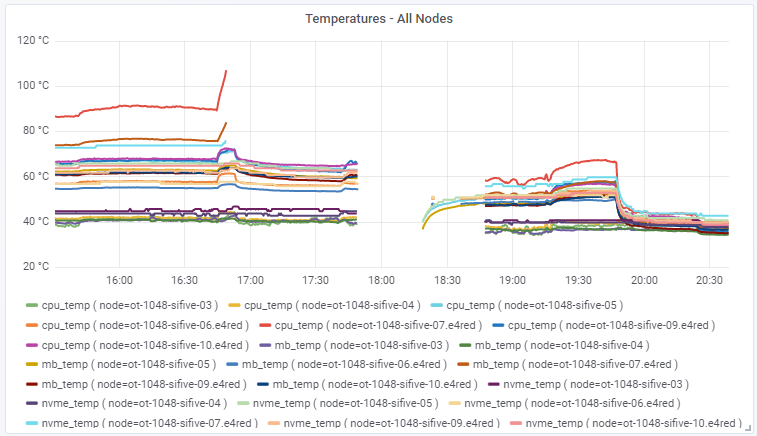}
    \caption{Thermal runaway during HPL execution}
    \label{fig:examon_thermal_runaway}
\end{figure}

    
Figure~\ref{fig:examon_thermal_runaway} reports the temperature measured at the available sampling points over time. We can notice that during the first HPL runs, we encountered a thermal hazard on node 7, which reached 107°C and stopped executing. We noticed that the nodes in the centre blades were significantly hotter than the remaining ones by further inspection. This is an effect of the 1U case and the suboptimal airflow design that needs improvements to remove the heat generated by the PSUs. We addressed this issue by removing the lid and increasing the vertical spacing between the blades. This led to a significant reduction in the hotter node temperature, from 71°C to 39°C.

\section{Conclusions and Future Works}  \label{sec:futureworks}

In this manuscript we presented \mcimone~:  To the best of our knowledge, this is the first RISC-V cluster which is fully operational and supports a baseline HPC software stack, proving the maturity of the RISC-V ISA and the first generation of commercially available RISC-V components. 
We also evaluated the support for Infiniband network adapters which are recognised by the system, but are not yet capable to support RDMA communication.

We characterised in detail the power consumption of the SiFive Freedom U740 SoC for different workloads, measuring $4.81W$ in idle,  with $64\%$ due to core power (32\% of leakage power, 51\% dynamic and clock tree power and 17\% by the O.S. workload), $13\%$ related to DDR and $23\%$ to the PCI subsystem. The power consumption increases to $5.935W$ under CPU intensive workloads.
Furthermore, we ported the ExaMon ODA system on \mcimone~and used it to detect thermal stability problems in the first configuration, which led to a thermal shutdown on the central node during the HPL run. We changed the enclosure to provide higher airflow to mitigate this issue.

Future work will focus on improving the software stack to achieve higher memory utilisation (i), to implement dynamic power and thermal management (ii), overcome the limitation in the Infiniband support (iv), extend \mcimone~with PCIe RISC-V based accelerators (v).

\section{Acknowledgments}
The study has been conducted in the context of the following projects: 
The european-project-initiative which has received funding from the European High Performance Computing Joint Undertaking (JU) under Framework Partnership Agreement No 800928 and Specific Grant Agreement No 101036168 (EPI SGA2). The JU receives support from the European Union’s Horizon 2020 research and innovation programme and from Croatia, France, Germany, Greece, Italy, Netherlands, Portugal, Spain, Sweden, and Switzerland.
The European PILOT project which has received funding from the European High-Performance Computing Joint Undertaking (JU) under grant agreement No.101034126. The JU receives support from the European Union’s Horizon 2020 research and innovation programme and Spain, Italy, Switzerland, Germany, France, Greece, Sweden, Croatia and Turkey.
The REGALE project which has received funding from the European High-Performance Computing Joint Undertaking (JU) under grant agreement No 956560. The JU receives support from the European Union’s Horizon 2020 research and innovation programme and Greece, Germany, France, Spain, Austria, Italy. 
The EPUEX project which has received funding from the European High-Performance Computing Joint Undertaking (JU) under grant agreement No 101033975. The JU receives support from the European Union’s Horizon 2020 research and innovation programme and France, Germany, Italy, Greece, United Kingdom, Czech Republic, Croatia.


\bibliographystyle{plain}
\bibliography{bib}

\end{document}